\documentclass{cta-author}

\usepackage[colorinlistoftodos, textwidth=65mm, shadow]{todonotes}

\usepackage{subfigure} 
\usepackage{multirow}
\usepackage{booktabs}

{}
{}
{}

\begin{document}

\supertitle{Brief Paper}

\title{Agile Risk Management for Multi-Cloud Software Development}

\author{\au{Victor Munt\'es-Mulero$^{1}$}, \au{Oscar Ripolles$^{1}$}, \au{Smrati Gupta$^{2}$}, \au{Jacek Dominiak$^{3}$}, \au{Eric Willeke$^{2}$}, \au{Peter Matthews$^{4}$}, \au{Bal\'azs Somosk\"{o}i$^{5}$}}

\address{\add{1}{CA Technologies, Strategic Research, 
%WTC Almeda Parc,
Pla\c{c}a de la Pau s/n, Edifici 2 Planta 4,
Cornell\`a de Llobregat, 08940 Spain\\}
\add{2}{CA Technologies, AgileOps,
3965 Freedom Circle, 
Santa Clara, CA 95054, United States of America\\}
\add{3}{CA Technologies, Strategic Research, 
ul. Woloska 5
02-675 Warszaw,
Poland\\}
\add{4}{CA Technologies, Strategic Research, 
Riding Court Road
Datchet, Berkshire
GB
SL39LL, United Kingdom\\}
\add{5}{Lufthansa Systems GmbH \& Co. KG, Siemensdamm 62
D-13627, Berlin, Germany\\}
\email{Victor.Muntes@ca.com}}

%\author{\au{S. Cheng$^{1,2}$}%%% First author
%\au{J.C. Ji$^1$}%%% Second author
%\au{J. Zhou$^2$}%%% Third author
%}

%\address{\add{1}{....}%%% Author address here
%%% First group represent author affiliation number and second group represent name.
%\add{2}{....}

\begin{abstract}
\looseness=-1
Industry in all sectors is experiencing a profound digital transformation that puts software at the core of their businesses. In order to react to continuously changing user requirements and dynamic markets, companies need to build robust workflows that allow them to increase their agility in order to remain competitive. This increasingly rapid transformation, especially in domains like IoT or Cloud computing, poses significant challenges to guarantee high quality software, since dynamism and agile short-term planning reduce the ability to detect and manage risks.
%In particular, agile methodologies have emerged to focus on quickly delivering quality Functional Requirements (FRs), as they usually represent FRs as User Stories. However, Non-Functional Requirements (NFRs) are usually not sufficiently identified, modeled, and linked and mechanisms to control risk related to NFRs are not clear. 
In this paper, we describe the main challenges related to managing risk in agile software development, building on the experience of more than 20 agile coaches operating continuously for 15 years with hundreds of teams in industries in all sectors. We also propose a framework to manage risks that considers those challenges and supports collaboration, agility, and continuous development. An implementation of that framework is then described in a tool that handles risks and mitigation actions associated with the development of multi-cloud applications. The methodology and the tool have been validated by a team of evaluators that were asked to consider its use in developing an urban smart mobility service and an airline flight scheduling system.

\vspace{0.5cm}
\emph{(This paper is a postprint of a paper submitted to and accepted for publication in IET Software (Vol. 13 , Iss. 3 , 6 2019) and is subject to Institution of Engineering and Technology Copyright. The copy of record is available at the IET Digital Library)}

\end{abstract}

\maketitle

\section{Introduction} \label{sec1}

Organizations in all industries recognize that they are continuing their evolution into technology and data companies \cite{McKendrick}, and that their business models are being partially or fully transformed by software. One of the main drivers of this transformation has been the adoption of paradigms such as IoT or Cloud computing. Cloud computing allows companies to create new business models that scale depending on the demand, to migrate their operations to the Cloud, and to create new service models such as Software as a Service (SaaS) \cite{mell}. Large technology companies acknowledge the importance of SaaS and Cloud computing\footnote{https://www.cio.com.au/article/619978/ca-technologies-realigns-saas-y-ux-driven-try-before-you-buy-market/}, and users also express their preference for a subscription-based model since customers become the focus of the companies instead of the product or the transaction\footnote{https://www.forbes.com/sites/kimberlywhitler/2016/01/17/a-new-business-trend-shifting-from-a-service-model-to-a-subscription-based-model/\#1517bc704a5f}.

%Cloud computing and IoT are two of the main enablers for industries to create new business models that scale depending on the demand of consumers. 
Gartner estimates that by 2022 the main barrier for IoT adoption will still be security~\cite{Gartner}. Northbridge indicated in their latest Future of Cloud Computing Survey \cite{Northbridge} that security was still the top concern and inhibitor of cloud adoption. Furthermore, Article 25 in the recent GDPR \cite{gdpr} discusses data protection by design and by default, underlining that considering privacy from the beginning is essential to address privacy successfully. Additionally there is an increased need to create trustworthy systems. Just as an example, Yan et al.~\cite{yan2014survey} published a survey where they describe the different dimensions of trust for Internet of Things systems. Their work concludes that risk management is an essential piece to guarantee trustworthiness.

Privacy or security-by-design can only be achieved if risk management is performed from the beginning of the software development cycle. Continuous and agile risk management processes are of great importance in IoT and Cloud computing since these environments have a complex and distributed network of services that increases the attack surface. In the context of multi-cloud applications, whose components are deployed over different infrastructures provided by different Cloud Service Providers (CSPs), risk related to accountability, assurance, agility or even financial aspects become even more challenging. Risks analysis can also guide the selection of CSPs, and several authors have proposed methodologies to consider risks in addition to quality of service or cost \cite{Omerovic}, especially in the context of multi-cloud applications \cite{gupta15}. Therefore, any component, whether running on premises, remotely hosted by a cloud service provider or offered as a service (by a CSP, a device in an IoT, etc.), is subject to risk analysis considerations. Poor risk management allied to a reactive strategy usually forces companies to continuously re-factor their application architectures to improve overall software quality and security, incurring in technical debt and high re-implementation costs~\cite{Boehm03}. 

The demands of software driven businesses and the need for fast innovation are forcing organisations to replace traditional software development methods such as Waterfall for %software development described by Royce~\cite{Royce87} and look for alternatives. 
alternatives that must be agile and support continuous development and delivery models in an aggressively changing market.
The principles that drive DevOps and Agile methodologies, with a focus on transparency and collaboration between teams, can help minimise the risks in the application and the inconsistencies in the way risks are managed. 
Nevertheless, the Agile Manifesto (agilemanifesto.org) suggests that Agile teams focus on delivering the simplest code that supports the needs of their customers. In their attempt to adhere to this bottom-up principle, Agile teams take an incremental and iterative approach that focuses on delivering individual software capabilities, usually derived from functional requirements (FRs), diminishing the relevance of system-wide quality attributes and other aspects related to non-functional requirements (NFRs), which in general are not well managed in agile methodologies~\cite{Paetsch03, Medeiros17}. This negatively affects code quality and the quality of the overall software architecture. 
As a consequence, elicitation and management of risks connected to NFRs in Agile become increasingly important to mitigate the negative effect of relegating NFRs to second-class citizens. %and the impact that this may have for coding, testing, and maintaining software.

To our knowledge, only some initial work has been done to conceptualize agile risk management techniques, and effective control of NFR-related risks in agile methodologies is still to be discussed. This paper aims at describing challenges that deserve more attention and then proposing and validating a solution. The paper is organized as follows. Section \ref{Sec_SotA} analyses how non-functional requirements are currently managed, and the main obstacles related specifically to risk management. This analysis will help us describe in Section \ref{Sec_MainChallenges} unsolved problems that pose challenges in agile risk management. As a solution to these challenges, in Section \ref{Sec_Methodology} we propose a framework to support a collaborative and agile risk analysis in environments that require continuous software delivery, in order to detect risks and generate sets of mitigation actions. Section \ref{Sec_Tool} describes how this framework and associated tool set have been implemented for the development of multi-cloud applications. This tool set was validated by a team of evaluators whose feedback is analyzed in Section \ref{Sec_Results}. Finally, Section \ref{Sec_Conclusions} draws some conclusions and outlines future lines of work.

\section{Related Work}
\label{Sec_SotA}

This Section discusses relevant work related to risk management methods for software development,
%both from a traditional perspective and from the point of view of Agile, where we also consider 
 non-functional requirements management in Agile methodologies, continuous software development, and risk management in a distributed agile development.

\subsection*{Traditional risk management for software development}
Implementing risk management processes is essential, specially for complex, high-risk projects \cite{Wallmuller:2002:RMS:643566.643578}. The business domain can also affect the need for risk analysis. In this sense, domains that are less volatile such as supply chain software will not change so quickly, while other businesses will be continuously looking at their customers and reacting to changes, modifying the requirements accordingly.

In general, previous work focuses on the classical schedule, budget and scope risk analysis. When using risk management techniques for software projects managed through waterfall methodologies, risks are thoroughly analyzed at the beginning of the project but when requirements change during the project life-time risk analysis may become obsolete~\cite{rao2016study}. Waterfall models assume requirements to be clearly defined in advance in the design stage and, in general, to remain fairly immutable, which tends to be unrealistic for projects in a rapidly changing market. Although there are change management mechanisms in waterfall processes that permit for requirements to change, risk management is seldom involved, and the overall risk analysis is not updated.

There are a number of risk assessment methodologies that include quantitative metrics, such as the probability of occurrence or the effort to implement control measures, or even its cost. But others consider qualitative metrics, e.g. an appraisal of project staff's motivation. There are many quantitative risk methodologies and tools, like RiskWatch\footnote{RiskWatch: https://www.riskwatch.com . Accessed: 2017-05-09} or ISRAM~\cite{Karabacak:2005:IIS:2625876.2626120} and there are many qualitative risk methodologies such as OCTAVE~\cite{Alberts02}, Coras~\cite{lund2010model} or one of the most commonly known in the hosted software arena, STRIDE \cite{STRIDE}. 
OWASP \cite{OWASP:2013} is an open standard trying to define the risk aware software development. It aims at making software security visible, so that organisations are able to make informed decisions. DREAD~\cite{Howard06} is a successor of STRIDE and provides another approach specialised in multi-stack type of applications. Pasha et al. \cite{Pasha} performed recently a detailed and thorough analysis of different risk management approaches, both for small and large-scale software systems. In addition to the state-of-the-art, the authors also proposed a methodology for risk mitigation that was very complete but lacks agility. 

\subsection*{Agile risk management for software development}

In general, it is accepted that managing risks and other non-functional requirements is ill-defined in agile approaches~\cite{Ramos18}. Different lightweight NFRs methodologies for agile processes have been presented such as ~\cite{Domah15} or \cite{Farid12}, where authors combined both FRs and NFRs in one framework and computed a risk-based requirements implementation sequence. Recently, Medeiros et al.~\cite{Medeiros17} propose an approach to specifying requirements based on design practices targeted to the developer. It is worth highlighting the work by Moran~\cite{Moran2014,Moran2014b}, where the author explicitly tackles issues related to risk management for agile software development. Moran proposes a risk modified kanban board and user story map. To our knowledge, this is the only research work that could be framed in the research challenges described in this paper. Some products like \emph{codeBeamer}\footnote{Risk Management in codeBeamer: https://intland.com/video/test-and-quality-assurance-management-videos/risk-management-in-codebeamer/}, which provide scaled agile capabilities, also include a mechanism to assign risks to both functional and non-functional requirements. However, their methods to define and manage risks are very similar to traditional risk management methods.

Most of the traditional risk management methodologies described above focus on the assessment of risks at a singular stage in time. However, these techniques do not address one of the most prominent challenges in today's adaptive, pivot-oriented world: the challenge of the continuously evolving risk profile. Managing continuous changes in software development has been studied from different perspectives. Practices such as \emph{``release early, release often''} have been promoted and adopted in open source software development \cite{feller05} 
and they prove to benefit software quality and consistency~\cite{Michlmayr15}. While continuous software integration~\cite{Stahl14} may be the most well-known practice in industry related to continuous software development, the increasing focus on security and privacy has also generated work on continuous security and regulatory compliance. Fitzgerald et al.~\cite{FitzgeraldS17} publishes a roadmap and agenda for continuous software engineering. Ameller et al.~\cite{ameller17} propose to re-plan the current release every time an activity triggers the need for an updated release plan.

Initially agile methods were devised to fit small projects with co-located developers in contexts where safety and security were not critical~\cite{Ambler01}. However, the use of these methods have been extended to large projects with distributed development teams \cite{Fitzgerald06} and safety-critical systems~\cite{fitzgerald13}. In this last paper, Fitzgerald et al. discuss R-Scrum (Regulated Scrum), an adaption of Scrum to support compliance in regulated environments, such as medical devices, railway, or aviation. They propose adding new ceremonies, artifacts and roles to allow compliance to be assessed at the end of each sprint.

Distributed agile development (DAD) approach is increasingly adopted by more and more software companies. The idea of DAD is to combine the quality and speed benefits of agile with the cost benefits of distributed software development (DSD). This combination generates significant risks, considering the contradicting nature of agile and DSD \cite{Mudumba10}. 
Shrivastava et al.~\cite{Shrivastava2017} present a risk management framework for DAD, studying risks related to software development life cycle, project management, group awareness, etc. However, this work is focused on analysing risk factors that represent a threat to the successful completion of a software development project, rather than risk threatening non-functional requirements. In \cite{Aslam2017}, the authors performed a systematic literature review on risks and control mechanisms for DSD, which could be extrapolated for a system to support DAD.

Finally, it is worth noticing that, in general, security tends to be considered lower in priority, sometimes unintentionally, when considered as one of the non-functional requirements of an application. Continuous security~\cite{Merkow11} aims at putting security as a high-priority concept through all phases of the software development process. Authors detect nine building blocks for continuous application software security, including for example employee training and awareness, creating a security software group or giving non-functional requirements the same importance as functional requirements. As an example, they suggest creating user stories to capture non-functional requirements. However, this does not provide a complete solution to challenges in Section~\ref{Sec_MainChallenges} since, for example, collaboration is not expressly addressed.

\section{Challenges in risk management for agile software development}
\label{Sec_MainChallenges}

In this section we present the main challenges for agile risk management related to software development. Before describing the challenges, it is worth discussing how risks are managed in Agile. The first wave of Agile adoption in many companies involved structuring software development through Scrum teams. %In general, companies find that aligning these teams against larger programs is difficult. Specifically, 
When it comes to application risk management, most companies cannot afford to have risk management experts in each team. A frequently used solution is to choose to have staff responsible for detecting the most important risks in a centralized way. The use of framework methodologies to scale Agile, such as SAFe\footnote{Scaled Agile Framework: http://www.scaledagileframework.com}, marginally improves this situation. In this case, organisations structure software development by creating teams of teams that plan and synchronise their work through \emph{Program Increments} (PI). %\footnote{ Program Increments: http://www.scaledagileframework.com/program-increment/}.
%and organise the work using a \emph{Program Backlog}\footnote{http://www.scaledagileframework.com/program-and-value-stream-backlogs/}.
When planning a new PI (typically every 10 weeks), all the actors participating in the software development process attend the meeting. 
%, usually in person. 
PI planning usually starts with a vision and a prioritized list of new features of the product%and the group comes up with the team and program PI objectives and the program board (description of feature deliveries, dependencies and milestones). During PI planning, features are
, which are then decomposed into \emph{user stories}. This is defined by product owners and created by the different Agile teams, which are in charge of drafting plans but also of analyzing risks and impediments. In fact, user stories replace traditional functional requirements, and describe intended system behavior. Consequently, risk analysis is done related to expected features and to these user stories. Because of this, non-functional requirements, which are usually not represented through these user stories, tend to be unintentionally diminished in terms of importance~\cite{Merkow11} and commonly ignored during the risk analysis.
During PI planning risks are identified by each team% over the first one-and-a-half days
, and then aggregated into a program risk sheet that is reviewed in a plenary session. The group then discusses and categorizes program risks and impediments. After PI planning, once risks have been classified, and actions and owners are established, all teams are assumed to be aware of these potential risks and impediments detected during the meeting and they are expected to act accordingly throughout the sprints in that PI. Additionally, the program leadership team will typically track those risks that are not resolved to ensure the right coordination occurs. Unfortunately, mapping between risks and user stories do not usually happen and, since PIs are guided by customer requirements, the risks detected remain as a relegated part of the development process. Risk management in an Agile environment must be integrated into the process, present in PI planning, in Sprint planning, and in any other ceremony. It must be a light but continuous activity.

The challenges included in this section are derived from the analysis of the related work and they are also based on CA Technologies Rally active coaching. %Tenths of thousands of people have been touched indirectly.
Our conclusions build on top of more than 20 agile coaches operating continuously for 15 years with ever-increasing scale of engagement. 
Authors of this paper accumulated more than 10 years of experience coaching companies to help them adopt Agile, between 2008 and 2017. We have directly interacted with more than 1,000 individual contributors. This includes a total reach of %approximately 55 release trains and
more than 8,000 people on hundreds of development teams spanning across six continents. Our conclusions are based on face-to-face coaching work in at least 9 countries and on the interaction with companies spanning all kinds of industries, from video game to health care to aeronautics to government, to give some examples.
As a result of these analysis, several challenges have been identified that need to be tackled to integrate risk analysis, and in particular risk analysis related to NFRs, in agile software development.

\vspace{0.2cm} 

\noindent \emph{\textbf{C1. Traditional risk analysis practices for software development do not easily translate to Agile.}}
Traditionally, %risk analysis tools %do not usually allow for efficient collaboration as required by a continuous delivery culture and they are designed for considering 
    risk analysis was performed "at design time" in the waterfall method, where design, development and operations are sequential and discrete worlds.
    These approaches usually rely on a risk management expert and it becomes impractical for small and medium enterprises in terms of cost and finding the proper resources. It also becomes impractical for companies working towards the adoption of more agile software development methodologies. 
    There exist some proposals for agile risk management~\cite{Moran2014}, but they do not take into account existing risk and threat analysis techniques such as STRIDE \cite{STRIDE} or DREAD~\cite{Howard06}, or other approaches for managing security risks like OCTAVE~\cite{Alberts02}, which are more focused on NFRs.

\vspace{0.2cm} 
    
\noindent \emph{\textbf{C2. Analysis of risks should be continuous.}} 
System and architectures evolve continuously. %, with the practice of collaborative \emph{emergent architecture} \footnote{Agile Architecture: http://www.scaledagileframework.com/agile-architecture/} being actively encouraged by the framework.
    In Agile, risk management is often neglected as part of a backlog or sprint plan, which leads to a view that it is not important until the later stages of a development project. In SAFe, PI planning sessions are useful to explore new risks that are foreseeable at the time of planning. However, a risk analysis every 3 months may not be sufficient when risks are detected by any actor at any time. %We lack tools for continuous analysis and evolution of risks between PI planning sessions.
	This exposes the general problem related to agile methodologies failing to effectively address risks and other non-functional requirements in a structured manner.
%\textbf{\emph{Enabling continuous risk management for highly dynamic systems that need to be continuously re-factored}}:
 There is a need for risk analysis methodologies that are adapted to agile contexts but still achieve the level of analysis and detail provided by \emph{traditional} risk assessment and mitigation techniques, in particular related to NFRs. Fitzgerald et al.~\cite{FitzgeraldS17} illustrate how Lean Thinking %~\cite{womack2003lean} 
    can be applied to continuous software engineering.
    However, they do not explicitly tackle challenges related to continuous risk management.
    Research into enabling continuous risk management can provide mechanisms and tools that resolve a challenge of implementing continuous analysis of risks. It is clear that a balance has to be struck between managing risk during the development process and not overloading or slowing the momentum of an agile methodology.

\vspace{0.2cm} 
    
 \noindent \emph{\textbf{C3. Teams do not have sufficient expertise on risk analysis.}}
 %Because delivering business value is usually not aligned with organisations working in functional silos (business, engineering, hardware, testing and Q\&A, etc), a
    Agile teams are usually cross-functional, optimised for communication and delivery of value. Although this could facilitate the work of specialised staff across several teams to analyse risks, allocating resources with expertise on risk assessment in each team becomes impractical. It is not possible to have a risk analysis expert in each agile team.
    This encourages some level of centralisation, but at the same time it requires transparency and for agile/Scrum team members, a higher capacity to participate and contribute in risk analysis and propose mitigation strategies when necessary. 
    %\textbf{\emph{Devising intelligent recommendation systems to mitigate the lack of expertise of different stakeholders}}: 
    Consequently, lack of expertise is an important issue for agile self-managed teams, but also for small or medium enterprises that cannot afford risk experts in each team. Detecting the most prominent risks and deciding the best mitigation actions may be a difficult task. Similarly, deciding the level of likelihood and impact of a particular risk may also be very subjective and, therefore, difficult to assess and measure. In addition, domain specific risk analysis further aggravates the situation as the level of expertise becomes more demanding.
    While training is an essential aspect \cite{Merkow11}, lack of expertise may always be present %in large and distributed teams
    and needs to be mitigated. %As the use of artificial intelligence becomes commonplace, an interesting research area emerges to create recommendation systems to help experts and inexperienced employees to define risks and the necessary mitigation actions. Besides, many tools to control software development allow %are made available to customers through SaaS offerings enables the use of novel crowdsourcing techniques to use the learnings extracted from the activity of past users in the platform. 

\vspace{0.2cm} 

\noindent \emph{\textbf{C4. Tools to manage risk in Agile do not foster collaboration.}}  
    %\textbf{\emph{Devising collaboration mechanisms and tools to enable collaborative risk management}}:  
    In order to improve the control of risk in Agile, tools that allow better transparency and enable collaboration of all stakeholders involved in the process are crucial. In this sense, creating new tools that can be easily embedded with other common agile tools to manage software development is very important. 
    Current tools to analyse and manage risks are quite limited. These are usually implemented through excel spreadsheets that are shared during PI planning, but do not allow for further collaboration beyond this face-to-face meeting. In general, there is a general lack of collaborative tools to engage all the stakeholders potentially involved in analysing risks in a transparent way, during PI planning and afterwards.
    Common generic tools such as Kanban style boards are increasingly used by software industry for scheduling work, representing user stories, features, etc. 
    In this sense, some works suggest the use of risk modified kanbans~\cite{Moran2014,Moran2014b}.
    %, or to extend well-established Scrum methods~\cite{Fitzgerald14}.     These tools need to be extended to integrate agile risk control methodologies. 
    The detection of new risks or planning mitigation actions should immediately propagate information and even trigger warning or actions through these software development management tools. 
    %Besides, the DevOps concept~\cite{%Debois09 
    %Humble11} emerged as an attempt to express the need for collaboration between development and operations that arise within software companies. %Collaborative risk management should also involve main actors in operations.
    There is an added challenge to handling risk in a collaborative way. Collective inter-team code ownership makes it difficult to control risks related to a particular component. In Agile, multiple teams frequently modify the code associated with a single component. Consequently, collective ownership makes it more difficult to control potential risks related to a particular component. 
    User stories will impact several components and many of them will typically need to modify the same shared component. Efficient tools do not exist to facilitate inter-team risk analysis. 

\vspace{0.2cm} 

Beyond these four challenges, it is also worth noting that there is also an intrinsic obstacle to improve risk management related to the organization culture. Cultural changes of any type require time and care to be successful. The challenge in this case is to develop a cultural change methodology that places risk management as a central and critical part of an agile methodology. 
    Cultural change of this type is often top down, implemented by management mandating risk management as an integral part of every sprint. This has the potential to lead to varying levels of adoption by the development staff. %CA has found in the past that significant changes to development methodologies such as a common development environment are adopted by three distinct groups of people, the innovators, the majority and the laggards. %These groups are based on the groups defined as part of the diffusion of innovation theory~\cite{rogers03} and adapted for use by CA internally by merging the early/late majority groups into one. 
    %The goal of a cultural change strategy is to increase the number of early adopters. %and majority to enable a focus on managing adoption by the laggard groups. 
    %More research is needed to develop a more general methodology for adoption of risk management at the early stages of a development life cycle.

\section{A New Framework for Agile Risk Management}
\label{Sec_Methodology}

In this section, we propose a new agile risk analysis framework to facilitate the creation of tools for agile risk management. This framework addresses the four challenges (C1-C4) described in Section~\ref{Sec_MainChallenges}. It is a framework that facilitates translating traditional risk analysis practices for software development to agile software development contexts, allowing for continuous risk analysis and permitting the main stakeholders, from the agile team members to any other level in SAFe, to collaborate together. 

This framework facilitates the control of risk associated to the \emph{assets} of a system. The materialization of an asset in each system may depend on the type of system and the main focus of the risk analysis. For instance, components in an application architecture of a software system being developed may be the assets to be analyzed. However, in a different situation, risks to be analyzed may be associated with user stories or product features. 

We conceive risk management as a continuous activity where risks may be subject to consideration and evaluation in the different stages of the software development life-cycle (C2). Further, the analysis assessment process is devised in a simple and visual way that simplifies the collaboration in a multi-disciplinary team of application development  stakeholders (C4), frequently referred to as the DevOps Team. We consider recommendations, constraints, and rules to guide and support the whole team and minimise the impact of having self-managed teams without adequate risk management capabilities (C3). Lastly, to address the agility challenge (C1) but also to foster continuity and collaboration, the Kanban process board philosophy is adopted to produce a risk assessment process board.    

Our framework uses a pull system in the style of Kanban, where the status of each asset with respect to a predefined risk analysis methodology is expressed through the different columns in the Kanban board. This makes our framework agnostic to any specific risk analysis methodology. A tool compliant with this framework links a Kanban-style representation with a particular risk analysis methodology, mapping each of the steps of that particular methodology to one of the columns in the Kanban. This naturally provides a solution to most of our challenges (C1, C2 and C4), as it provides a mechanism to include traditional risk analysis methodologies into an agile-ready tool such as the Kanban. Kanban style boards are used in continuous deployment environments, fostering collaboration among team members and among teams. %In particular, it naturally allows for collaboration through well-known software development agile methods and a Kanban style interface and 
It provides an intuitive visual representation of the current status of your system with respect to risks/threats and the implemented mitigation strategies. It also allows introducing new risks and threats and re-evaluating them in relation to each system asset,  aligning with continuous software delivery approaches.

\subsection{Input data assumptions}
The framework presented in this paper considers that risk assessment will rely on three essential sources of information:
\begin{itemize}
\item The definition of the system/application's assets. For example, these assets can be components of the application architecture, components in the infrastructure of a physical system, or user stories or features of a particular software product. 

\item The set of constraints or rules that the team has selected. As examples, these rules may be used to prevent moving an asset until some restriction has been met, to decide when warnings or specific messages should be shown, or to indicate that an approval from some member of the team must be obtained before performing an action. These rules will enforce the validity of the risk analysis and guide non-experienced users.

\item The knowledge database. This database contains information about types of assets, risks, and mitigation actions, as well as their relationships. This information may be used to automate  some processes and provide recommendations that will guide the team through the different steps of the selected risk analysis methodology. This knowledge database may be adapted to the domain of the application to be developed and extended by the users.

\end{itemize}

\subsection{Mapping of existing risk analysis methodologies}
Our framework proposal assumes that methodologies for risk analysis can be divided into steps that can be implemented sequentially. Let us define $S=\{s_1,s_2,...,s_n\}$ as the set of $n$ steps of a given methodology. Our framework involves defining a set $C$ of columns in the Kanban-like board with $n$ columns ($C=\{c_1,c_2,...,c_n\}$). With these two sets we can create a mapping so that each column of the Kanban board refers to each step in the methodology. For example, OWASP \cite{OWASP:2013} proposes 5 steps for rating risks: identify risks, estimate likelihood, estimate impact, determine severity of risk, and decide what to fix. We would then have 5 different columns in our Kanban.

\begin{figure}[!htb]
\centering{\includegraphics[width=\linewidth]{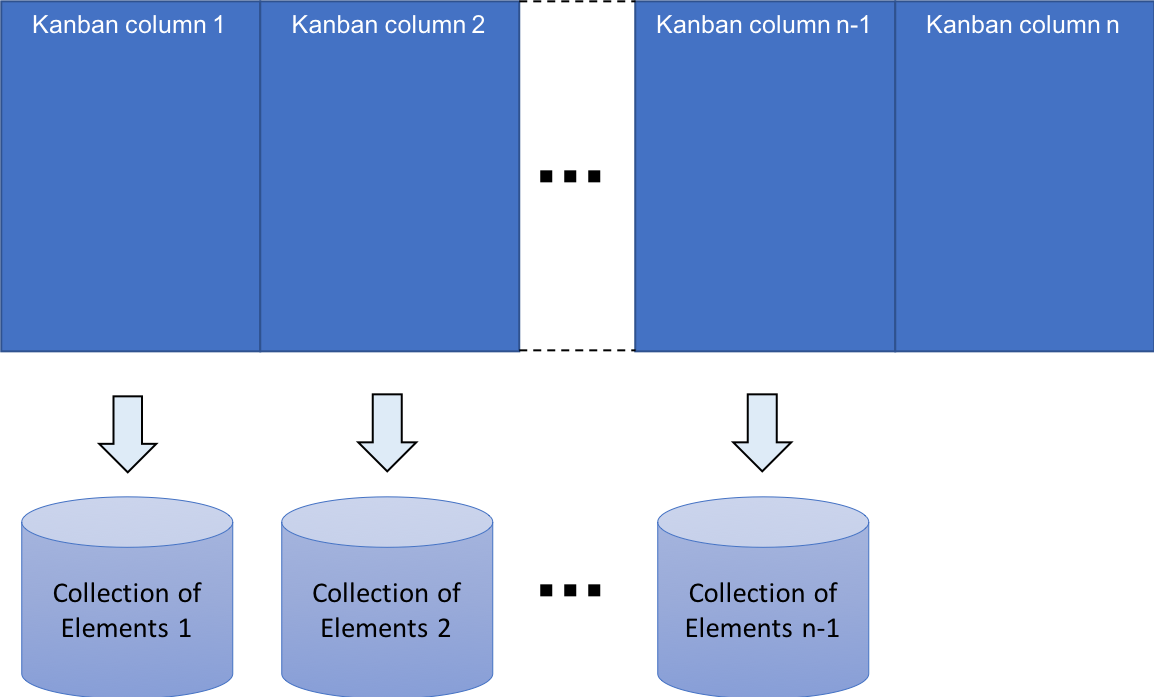}}
\caption{Kanban-like board with $n$ columns that stores in a storage system a ``Collection of Elements $x$'' generated in each Kanban column $x$. \label{FigKanbanColumns}}
\end{figure}

Figure~\ref{FigKanbanColumns} depicts a generic representation of the Kanban approach proposed in this framework, where each Kanban column generates a collection of elements. These collections of elements are stored in a storage system. The actual content and semantics of these data will depend on the selected risk analysis methodology. For instance, a column may store collections of vulnerabilities, threats, risks or mitigation actions, to give some examples.

\subsection{Agile Risk Analysis Automation}

As we mentioned before, one of the aims of this framework is to reduce the impact of having teams with little risk-related expertise by adding some automation to the framework. This allows team members, who may not be experts in risk analysis, to participate in the implementation of risk analysis methodologies with the support of the system, thus tackling the challenge depicted in C3. We propose solving this challenge from two perspectives. One the one hand, and depending on the risk methodology selected, some columns may rely on a recommendation system to support the stakeholders. In this sense, recommendation of vulnerabilities, threats, risks or mitigation actions may be provided. In multi-tenant environments, recommendation may be based on the anonymized information collected from other users using a particular tool. All this information will be available in the knowledge database mentioned before.

On the other hand, our framework defines a set of constraints over the movements of the assets represented in the kanban-like board. These constraints depend on the risk analysis process in the background linked to the board. We propose defining a set of rules $R=\{r_1,r_2,...,r_m\}$ that constrains what type of actions can be done in a particular step depending on the methodology chosen. Any tool compliant with this framework should include a \emph{Movement Approval Module}. When a component is moved from one origin column in the Kanban-like board to another target column in the board a query is generated in order to evaluate a condition relating elements of the origin column to the elements generated in previous columns. 

\begin{figure}[!htb]
\centering{\includegraphics[width=\linewidth]{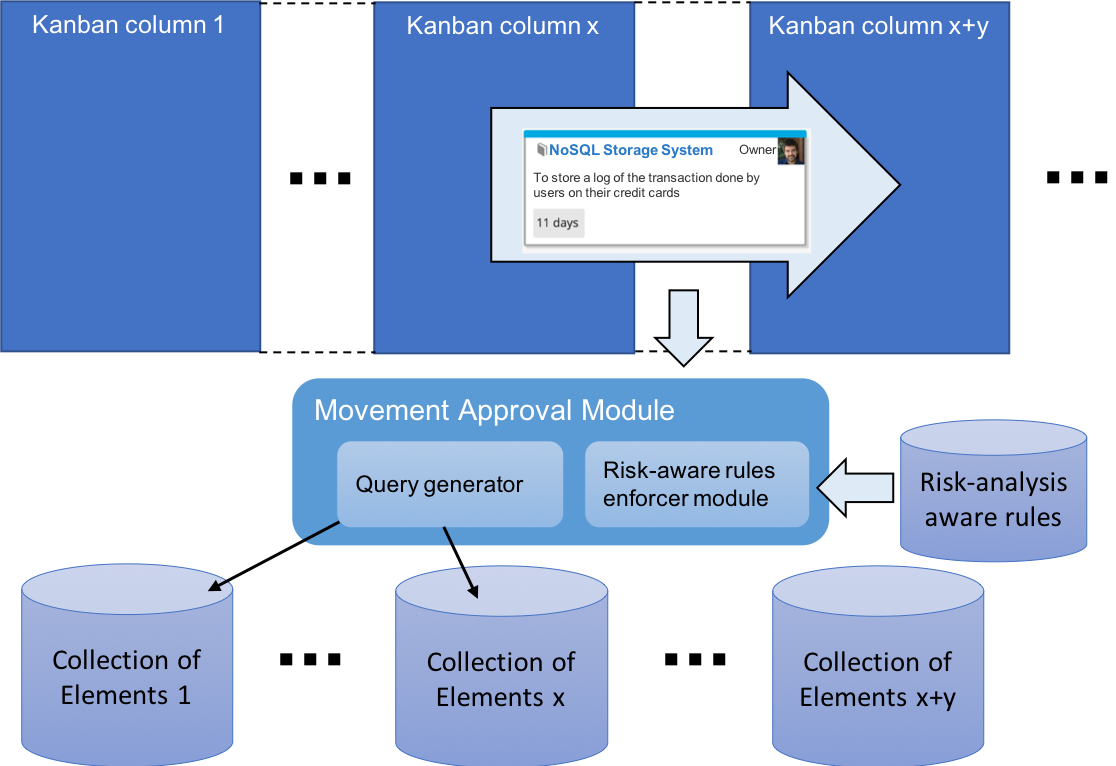}}
\caption{Movement Approval Module is activated by each movement of a component between Kanban columns. This module checks conditions and allows or disallows. \label{FigKanbanControlSystem}}
\end{figure}

Figure~\ref{FigKanbanControlSystem} describes this process. The movement of a component generates a call to the Movement Approval Module. This module is in charge of accepting or rejecting a drag and drop action. In order to do this, this module takes into account a set of rules that impose restrictions on the risk analysis process. A couple of examples of these rules may be:
\begin{itemize}
    \item A component cannot be moved from column $c_i$ to a column $c_{i+x}$ where $x>1$.
    \item A component cannot be moved from column $c_i$ to a column $c_{i+x}$ where $x>0$, if there exists an element generated in column $c_{i-1}$ that does not have an element generated in column $c_i$ linked to it (e.g. you cannot move a component to the mitigation actions phase if there are threats that do not have risks associated to them while the component was in the risk definition column)
\end{itemize}

The query generator will generate queries on the storage system to collect the data necessary to validate the conditions imposed by those rules.

When a movement is rejected, the tool may generate different types of feedback including the automatic return of the component to the origin column, a warning message providing explanations that justify why the movement is not legal in that context, a warning icon in the component that provides a justification upon click, etc.

%The we could say that our framework assumes that each column has a dataset of elements associated. The meaning of elements could vary depending on the meaning of the column and the particular step. 

\subsection{Additional Aspects}

A tool implementing this framework may allow marking some of the elements generated in a column as ``deferred''. When an element stored in the system is marked as deferred, it means that it can be omitted by the Movement Approval Module. For instance, a vulnerability of a component is detected but the architect knows that this vulnerability will not be important during the first year of the project. Its analysis can be deferred. Marking the vulnerability as deferred, the system would allow the movement of that particular component with that vulnerability to the following column without blocking the risk analysis of that component.

The proposed framework also allows for tools to include additional support forms to prepare a component to be moved to the next column. These forms may be different and generated ad-hoc depending on the current column of each component and the semantics of that particular column in the mapped risk analysis methodology. For instance, for risk definition we may provide forms to connect each component to risks and scores for evaluating those risks. 

\section{Risk Assessment for Multi-cloud Applications}
\label{Sec_Tool}

\begin{figure*}[!htb]
\centering{\includegraphics[width=15cm]{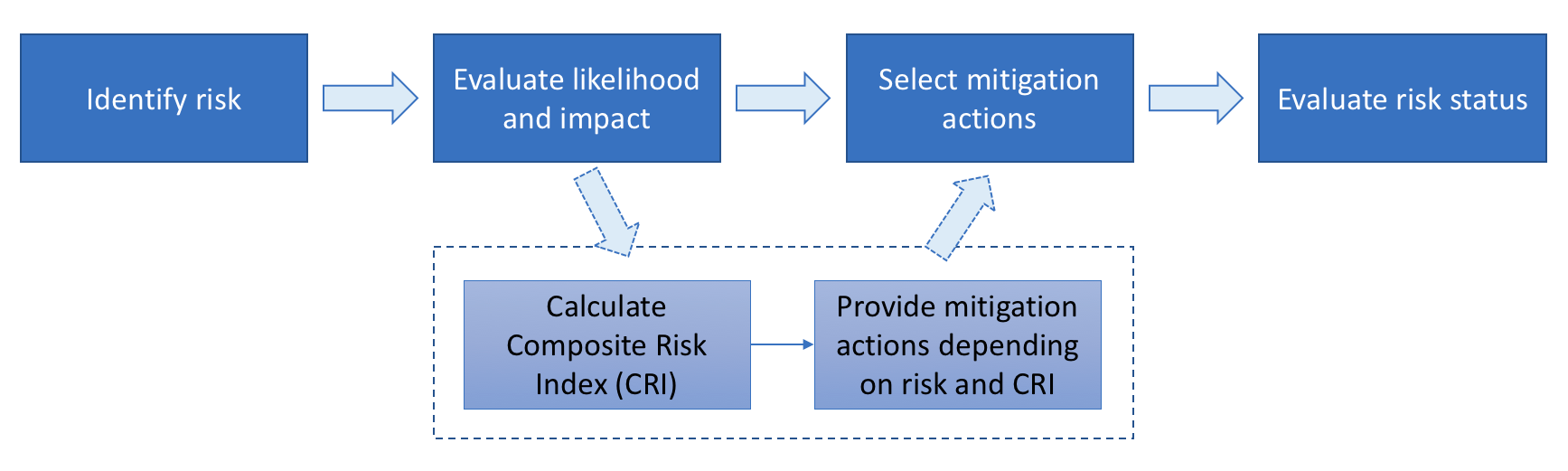}}
\caption{Outline of risk assessment process flow. Steps inside the dotted box are performed automatically.\label{fig1}}
\end{figure*}

In order to evaluate the proposed framework, we  developed a tool to support a risk assessment methodology for multi-cloud application development. A multi-cloud application distributes its components over heterogeneous cloud resources but, from the user's perspective, it works in an integrated and transparent way. Risk associated with these applications include those of its individual components but also those related to the overall security and to the data communication among its components. In this Section, we will first detail the risk methodology that we have selected for our tool and then describe how it was implemented. In any case, it is worth noting that the risk  methodology is explained for illustrative purposes as an example. Our agile risk management framework is agnostic to a particular methodology and other examples with different steps could be mapped in it.

\subsection{Selected methodology for the tool}

Figure \ref{fig1} outlines the flow of the selected Risk Assessment methodology, which is compliant with our proposed framework and has been inspired in the OCTAVE methodology \cite{Alberts02}. The selected methodology is composed of four main steps to be followed when the team wants to perform the risk assessment of one of the components:

\textbf{Risk identification.} 
As mentioned above, a knowledge database is used to support the user while identifying the risks. Depending on the type of asset, a subset of the possible risks is shown. The user is then asked to select one or more risks from the knowledge database. The user is also allowed to add new risks to the database if the ones suggested do not cover the specificities of the asset being analyzed.

\textbf{Risk evaluation.} 
For each risk that the user has selected, an evaluation of the likelihood and potential impact that the risk may have is performed. To evaluate each of the selected risks, the user is required to provide the likelihood and impact of each threat. With this information, the Composite Risk Index (CRI) \cite{Banerjee} of the risk is evaluated following equation \ref{eq1}. Both likelihood and impact are computed on a scale of 1-9 and the product is quantised on the scale of 1-5. This implies that the CRI ranges from 1 to 25.

\begin{equation}\label{eq1}
CRI=Likelihood * Impact
\end{equation}

%To support the team in this task, our methodology also considers the possibility of storing in the knowledge database a set of pre-calculated likelihood and impact values per risk, as well as a set of influencers for both. These influencers may be simpler to evaluate and, combined together, would offer the final values for likelihood and impact.

\textbf{Mitigation actions selection.} 
This step allows the user to discover the means to mitigate each of the risks. After evaluating the risk scores, risks are categorised according to their CRI level as those requiring treatment (high and medium risk level) and those that may not require treatment (low risk level). The knowledge database will present the most probable mitigation actions, but the user is free to add any other action from the knowledge base.  

\textbf{ Risk status evaluation.} 
Once the selection of the security controls of a risk is complete, the user is asked to select its ROAM status. ROAM \cite{Baah} is a common agile management risk mitigation classification whose acronym stands for: 
    \begin{itemize}
    \item Resolved - the risk has been answered and avoided or eliminated.
    \item Owned - the risk has been allocated to someone who has responsibility for doing something about it.
    \item Accepted - the risk has been accepted and it has been agreed that nothing will be done about it.
    \item Mitigated - action has been taken so the risk has been mitigated, either reducing the likelihood or reducing the impact.
    \end{itemize}
It is important to note that only risks with status Accepted or Mitigated are considered as fully addressed. Status Owned is treated as a pending status therefore the risk mitigation analysis must continue and Resolved status eliminates the prior risk analysis all together since the threat is considered no longer relevant.

%Finally, we should keep in mind that we should repeat this process for any possible risk that we may detect for a given component, and that this process should be iterated as many times as needed by the own iterations of the development process.

\subsection{Implementation of the tool}

Once the methodology was clear, a tool was developed to support the development of multi-cloud applications. This tool was also developed to address the prior challenges identified.

As we described in the previous section, our risk assessment framework relies on three sources of information:
\begin{itemize}
\item The definition of the application assets: in this case, the assets that the tool will consider will be components of the architecture of the application. Components may range from small components of the architecture in the form of specific purpose libraries running on premises to complex and general components in the architecture including sub-components or complex services offered by cloud service providers, by devices in an IoT ecosystem, etc. 
In the area of cloud applications, there have been many attempts at defining a domain-specific language (DSL) that can describe cloud applications \cite{Moran, CloudView, CloudML}. It is worth mentioning that the OASIS technical committee called TOSCA (Topology and Orchestration Specification for Cloud Applications) is developing an open standard that provides a language to describe cloud components and their relationships \cite{TOSCA}.
In our case we have chosen CAMEL (Cloud Application Modelling and Execution Language) \cite{CAMEL}, a DSL akin to TOSCA that allows users to specify multiple aspects of cross-cloud applications, such as provisioning and deployment, service-level objectives, metrics, scalability rules, providers, %organisations, users, roles, 
security controls, execution contexts, and execution histories. Using CAMEL, the development team is able to describe the architecture and the deployment requirements with a high-level of abstraction and independently of any cloud provider.

\item The set of rules that the team has selected: for simplicity, we have not established roles and every user has the same responsibility over the risk assessment process. The rules that have been incorporated to the tool are: (i) A component cannot be moved from column $c_i$ to a column $c_{i+x}$ where $x>1$; (ii) a component cannot be moved to the Mitigation actions selection unless all the risks have been evaluated and their CRI calculated; (iii) a component cannot be moved to the Evaluation column unless all the risks have at least one security control; and (iv) the risk analysis of a component cannot be considered as fully addressed unless all the risks have been accepted or mitigated.

\item The knowledge database: in order to assess the risks, in our tool we use a risk model based on the OWASP risk modelling \cite{OWASP}
and we gather information from different sources, such as the OWASP TOP 10 threats catalogue \cite{OWASPTop10} or NIST SP 800-53 r4 \cite{NIST}. This knowledge database is based on a predefined set of possible risks (here called threats) and a matching set of mitigation actions (here called security controls) which needs to be fulfilled by the application designer. Each of these security controls come with the definition and measuring technique on how the security control should be fulfilled.
\end{itemize}

%Our framework receives as input the application architecture, which describes the different components of the application up to the level of granularity desired by the user. The architecture definition may be done in any format, for instance, an UML diagram stored in xml format or standards such as the ISO/IEC/IEEE 42010:2011 \cite{ISO}.

Once the architecture is ready, the components to be analyzed from a risk perspective are imported into the Kanban. It is important to note that we have added an initial state for those components for which we have not started analyzing risks and a final state for those components for which the risk analysis is finished. Also, we have decided to combine the four steps of our methodology into two steps: identification and evaluation of risks, and selection and evaluation of mitigation actions. Thus, the four states and columns that our Kanban offers are:
\begin{itemize}
\item Components definition, which is the initial step for all the components pending their risk assessment.
\item Risks definition, where the users would move the components to start the risk assessment. In this step, the users are asked to decide the risks that affect the component. Moreover, in this step the users are also asked to evaluate the likelihood and impact of the risks.
\item Security controls definition, where the users are presented with the possible security controls of each risk depending on the CRI. Once the users select the security controls, they are also asked to apply ROAM to the risks.
\item Validation, for all the components that have finalized the risk assessment. Only those components whose risks have been Accepted or Mitigated should be in this state. The last step required from the user is the acceptance of the level of the risk mitigation status. The Validation step provides an overview summary of the choices made in previous steps.
\end{itemize}

%Figures \ref{fig2} to \ref{fig5} show how components are handled in our tool while they go through the different steps. In the remaining of this Section we will detail the steps regarding the risk and the security controls definition since they are the core of the risk analysis process.

%\begin{figure*}[b]
%\centering{\includegraphics[width=17cm]{Picture2.png}}
%\caption{Risk Assessment Process Board \label{fig2}}
%\end{figure*}

\subsubsection{Risks definition}

In this step, the user chooses the threats that the component under consideration is susceptible to.
%These threats are part of the knowledge database in the form of a threat catalogue that includes potential threats taken from a number of sources such as the OWASP TOP 10 threats catalogue \cite{OWASPTop10}. % The user is provided the flexibility to change the default values of the threat factors at every stage of risk assessment. 
Once threats are selected, they are automatically classified in the STRIDE \cite{STRIDE} categories (Spoofing identity, Tampering, Repudiation, Information disclosure, Denial of service and Elevation of privilege).

Regarding the evaluation of the risks, the likelihood and consequence scales chosen are inspired from \cite{STRIDE}. 
For simplification, CRI is also provided as an option for the user to provide likelihood and impact for each of the STRIDE categories, and the same scores are applied to all the threats categorised under each of the 6 categories of STRIDE). 
In our risk assessment process, the Likelihood and Impact values are further computed from a set of categorisations-based influencers taken from OWASP approach to the CRI. These influencers simplify the process and include concepts both from a technical perspective (ease of exploit, skill level of the threat agents, etc.) and the business perspective (financial damage, reputation damage, etc.). These sub-values are grouped by the type of factors and represented by the value in a scale of 0-9 where 0 represents a very unlikely scenario and in contrast 9 represents a very high likelihood of the factor to occur. Detailed description of all the factors can be found in \cite{OWASP}. Most of the Impact factors are pre-populated with values based on our threat catalogue. %In our specific implementation, business impact factors were not set to default values due to their dependency on the assessed business in question. Therefore, the user will always need to set the values of the business impact factors.

\begin{figure*}[b]
\centering{\includegraphics[width=17cm]{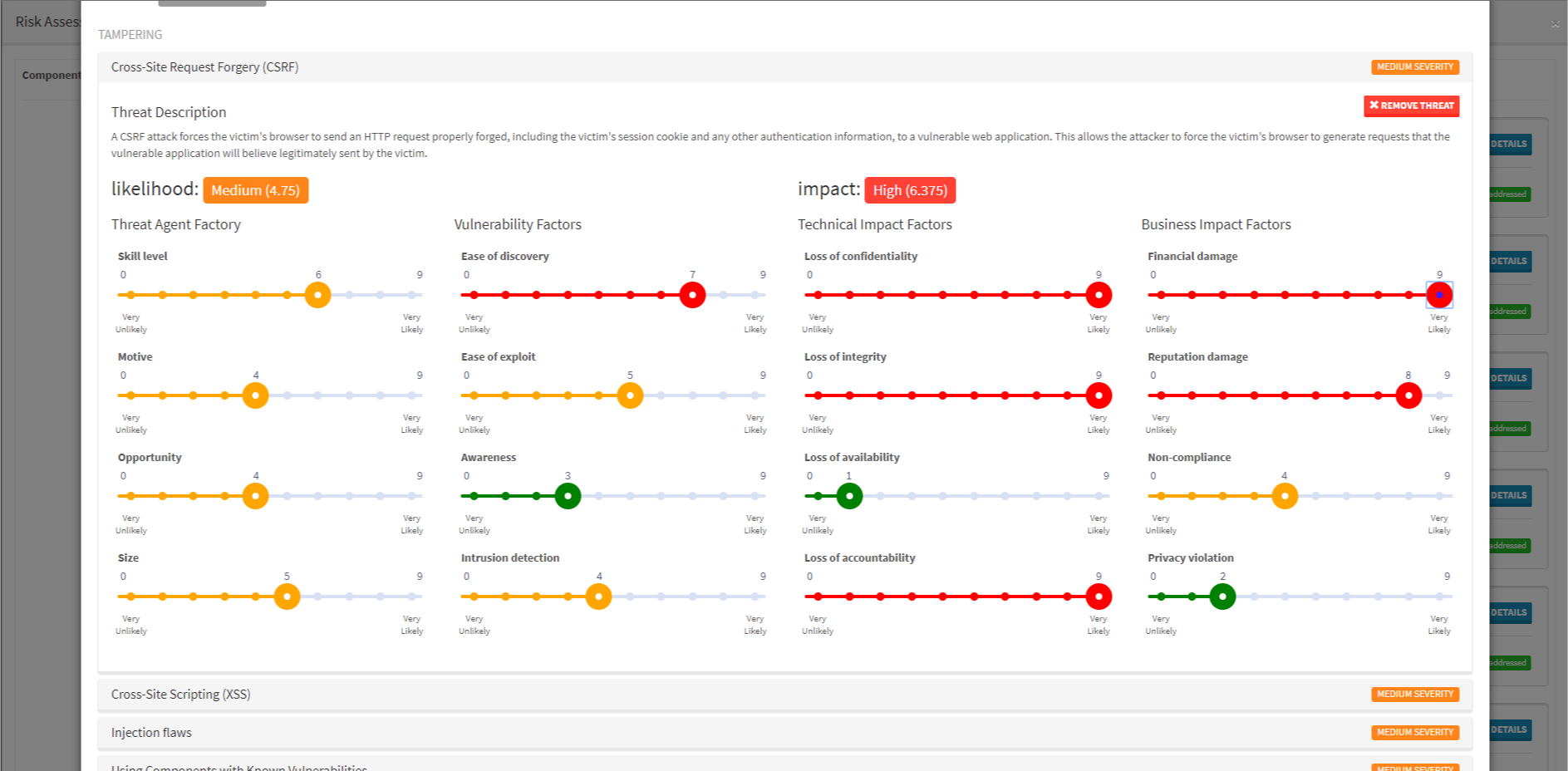}}
\caption{STRIDE based Risk Assessment step \label{fig3}}
\end{figure*}

In our tool, selecting the Details button for a component in the STRIDE based Risk Assessment column would bring up the screen in Figure \ref{fig3}. This shows the STRIDE Risk Assessment process and the actions that need to be followed to handle this assessment. Figure \ref{fig3} shows several rows representing the STRIDE defined threat categories, for example Tampering, Information Disclosure, Denial of Service. It also shows the likelihood and impact specification using the OWASP guidelines to compute likelihood and impact based on Threat Agent factors, Vulnerability factors and both technical and business impact factors. As many of these areas that are relevant have been completed, the user can move to the next stage of the risk assessment.

\subsubsection{Security controls definition}

Within the cloud security arena, this can be done by selecting the security controls the provider needs to guarantee in order to mitigate the threat. As indicated before, NIST SP 800-53 r4 \cite{NIST} maps security controls to the threats and indicates the threat levels that require treatments. Based on this mapping, the required controls are obtained for the threats selected by the users. These controls are then presented to the user as suggestions but, as mentioned before, the user is free to extend the choice to all the available security controls if desired. Selected controls are further mapped to the CCM (Cloud Control Matrix) controls from Cloud Security Alliance (CSA) \cite{CSA}.

\begin{figure*}[b]
\centering{\includegraphics[width=17cm]{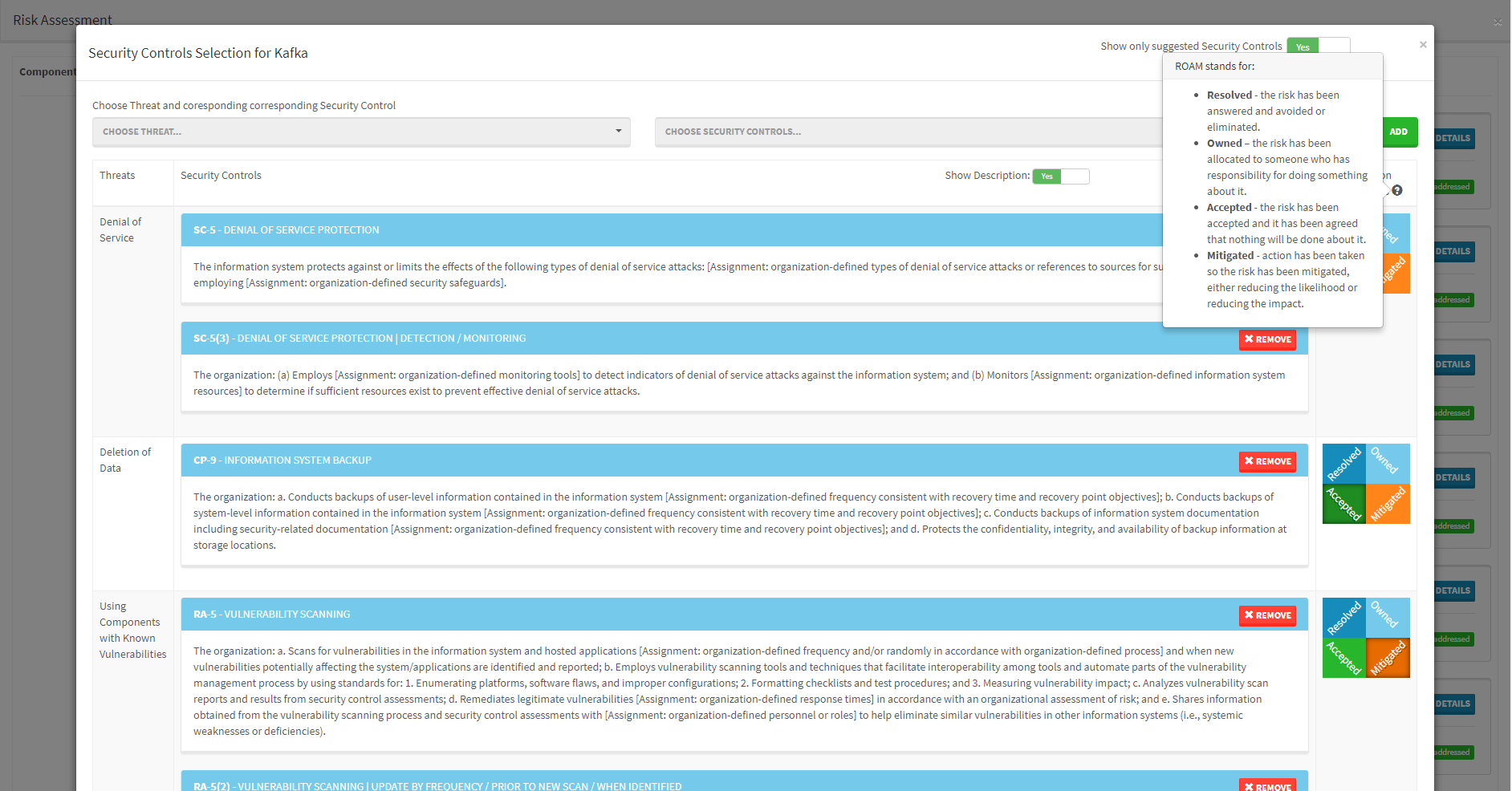}}
\caption{Security Controls Definition step \label{fig4}}
\end{figure*}

Figure \ref{fig4} shows the contents of a component that is in this stage. As we can see, in addition to the security controls, the user is also asked to apply ROAM to the risks.
 
We should keep in mind that we should repeat the whole risk analysis process for any possible risk that we may detect for a given component, and that this process should be iterated as many times as dictated by the iterations of the development process. Consequently, components that are in the Validation state may be moved to previous states if necessary. It is  important to note that in any state, the user can request the tool to output a report detailing the risk assessment status of each component, including the risks identified, their status, and their security controls. 

%\begin{figure*}[b]
%\centering{\includegraphics[width=17cm]{Picture5.png}}
%\caption{Validation step for a fully performed risk analysis on a sample component. \label{fig5}}
%\end{figure*}

\section{Results}
\label{Sec_Results}

Research into adoption of risk management as an integral part of Agile requires both qualitative and quantitative analysis. %It is difficult to gain accurate figures for survey responses as the dependencies, target audience, quality of survey, method of surveys have a major influence. Survey has low levels of response, with web based surveys coming in generally at approximately 11 percent and email at approximately 18\% \footnote{FRYREAR, Andrea 2015: Survey Response Rates: SURVEYGizmo  https://www.surveygizmo.com/survey-blog/survey-response-rates/ Accessed May 2017}. This can be potentially increased in an internal company survey to between 30 and 40\% which delivers more meaningful results. 
A survey can help us gather quantitative figures while a qualitative approach using an intensive case study is an effective way of understanding a team's approach. Case studies not only deliver insight into the thinking and ideas of a development team but allows their actions, behaviour and body language to be observed and recorded for analysis. %The case study approach enables observable behaviours and attitudes to modify adoption plans and time tables, or modify the methodology for adoption.
A combination of both approaches compliment each other and will facilitate further analysis and development of relevant hypotheses. To address this objective, we will firstly introduce the case studies and analyze the selected team of evaluators before presenting the results of the evaluation obtained through surveys. Then, we will also discuss the opinions of the evaluators regarding how relevant risk management is when compared to other tools used to develop multi-cloud applications.

\subsection{Use cases description}

For this evaluation, two different real case studies were chosen: an urban smart mobility service and an airline flight scheduling system. 

The smart mobility application should provide efficient and optimal route planning by considering road, traffic, energy consumption, and weather conditions.
The urban mobility service was proposed to have 4 components: the smart mobility engine that would serve as orchestrator, the consumption estimator that would calculated the energy needed on each trip, the multi-modal journey planner that would offer the optimal trip, and the database.

Airline scheduling is a complex scenario since each airline must react to actions of the rest in order to keep the schedules up to date. The flight scheduling system was proposed as an application with 5 components: the central gateway that would serve as entry point, the read module that would query fleet and airline-related information, the write module that would update fleet and airline-related information, the web interface that the final user would interact with, and a set of additional cloud services, such as event managers or databases.

In both use cases the objective was to provide a distributed solution that could reduce the points of failure and offer greater flexibility. Moreover, these application may also become Platform-as-a-service solutions that could scale or do load balancing as needed. With this plan, a user-centered evaluation was performed to assess if the tool developed fulfilled the needs of the users.
%The urban mobility service was proposed to have 4 components:
%\begin{itemize}
%    \item the smart mobility engine that would serve as orchestrator,
%    \item the consumption estimator that would calculated the energy needed on each trip,
%    \item the multi-modal journey planner that would offer the optimal trip,
%    \item and the database.
%\end{itemize}

%The flight scheduling system was proposed as an application with 5 components:
%\begin{itemize}
%    \item the central gateway that would serve as entry point,
%    \item the read module that would query fleet and airline-related information,
%    \item the write module that would update fleet and airline-related information,
%    \item the web interface that the final user would interact with,
%    \item and a set of additional cloud services, such as event managers or databases.
%\end{itemize}

\subsection{Evaluators team}

\begin{figure}%
\centering
\subfigure[Distribution of job positions.]{%
\label{fig:first}%
\includegraphics[height=2in]{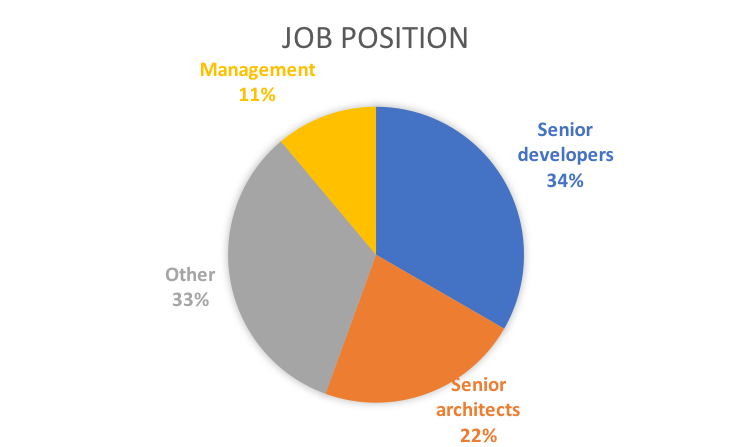}}
\qquad
\subfigure[Familiarity with cloud-based application development.]{%
\label{fig:second}%
\includegraphics[height=2in]{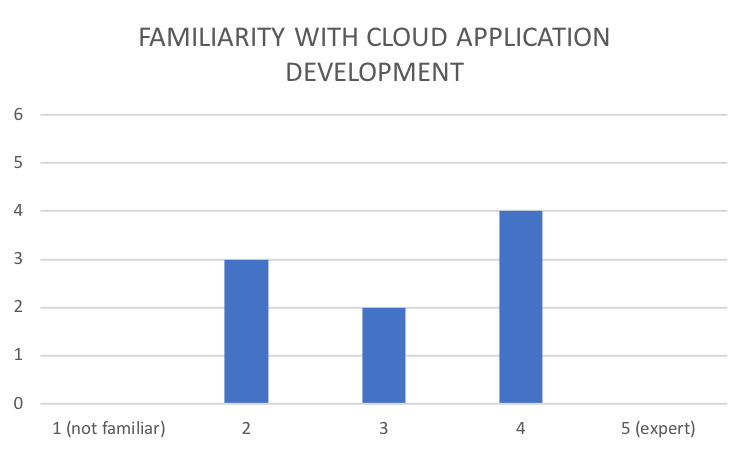}}
\subfigure[Familiarity with risk assessment.]{%
\label{fig:third}%
\includegraphics[height=2in]{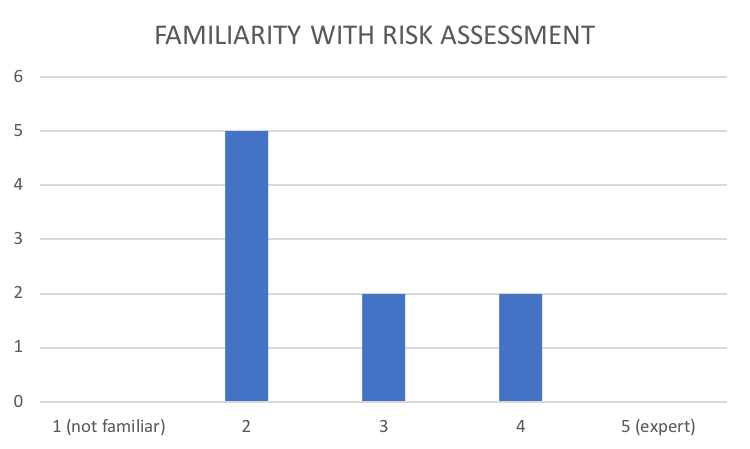}}
\caption{Analysis of the group of evaluators.}
\label{fig:evaluators}%
\end{figure}

To evaluate our methodology and our tool, we selected a group of evaluators from Lufthansa Systems and Tampere University of Technology. We were looking for evaluators that were not familiar with the tool, who were experienced defining architectures with CAMEL, who had a limited level of expertise on security, and who covered different levels of expertise in cloud development. Figure \ref{fig:evaluators} shows different information of the 9 evaluators. The figure shows how the job positions are distributed, what was their relation to cloud-based application development, and how familiar they were with risk assessment. We can see how most of the evaluators were partly familiar or very familiar with the development of cloud applications, whereas most of them were only slightly familiar with risk assessment. In this sense, according to evaluators's experience, risk management followed a quite rudimentary approach, without a formal or systematic approach to evaluate risk and tackle attack vectors. In the case of Lufthansa Systems, previous developments were for internal use and risks were not a priority. In general, risks were managed during the early planning phases (and in most of the cases only once) as it was considered as a one-time activity rather than a continuous process. The applied tool for this exercise was usually MS Excel, and in most of the cases there was only one person responsible for filling out the sheets with mostly NFR related risks. For Tampere University of Technology, focus was put on solving the functional requirements of the application. When having meetings with potential customers, they gathered their security concerns as a checklist of non-functional requirements. Then these were ordered by priority and tackled with security controls. In this sense, risk management was mostly done on the fly.

Finally, and although not shown in those figures, it is also worth noting that all the evaluators considered themselves as having a good knowledge of Agile.

%The tool has been analyzed from 3 quality perspectives:
%\begin{itemize}
%\item Efficiency, understood as a measurable concept that relates the output obtained to the input used. If efficiency %increases, the amount of waste or unnecessary efforts decreases.
%\item Usability, understood as the ease of use or of learn.
%\item Flexibility, understood as the ability of the tool to adapt in response to external input.
%\end{itemize}

%In the remaining of this section, we will consider each of these perspectives separately, as well as some other remarks and results obtained during the evaluation. Nevertheless, we will start by detailing the recruitment process of the evaluation team.

%\subsection{Efficiency}

\subsection{Use cases evaluation results}

For the evaluation we divided the evaluation group into 2 teams. Each team was asked to consider a different use case and then we collected their feedback in the form of a survey. 

\begin{figure*}[!htb]
\centering{\includegraphics[width=17cm]{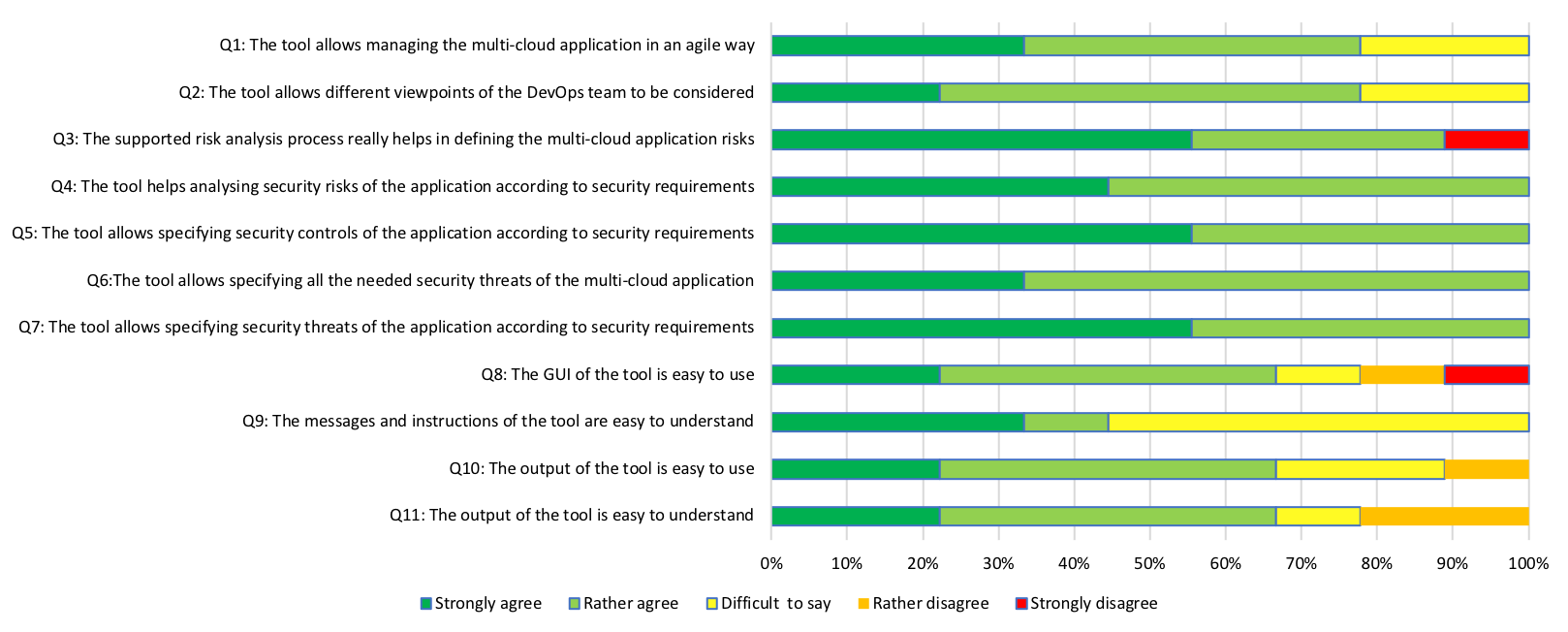}}
\caption{Results of the evaluation.}
\label{fig:efficiency}
\end{figure*}

Figure \ref{fig:efficiency} presents the answers given by the evaluators to different questions about the proposed risk analysis tool. According to the evaluators, the tool supports DevOps collaboration and it is efficient in the security risks definition aligned with the application security requirements. We can also see how the majority of the evaluators agree that the tool supports the agile management of multi-cloud applications. This shows that our proposed tool can help mitigate three of the challenges (C1, C2, and C4) presented above related to the lack of agile tools, the need for continuous risk assessment, and to the lack of collaboration. Given that the evaluators consider the tool to be easy to use and the supported process to help the risk analysis process, we also consider that our tool helps minimizing challenge C3, since it can help alleviate the lack of security expertise in the teams. The evaluators considered that the output of the tool is easy to understand but although the tool is easy to use, they believe that the messages and tooltips shown were sometimes confusing. We consider that some future work could be dedicated here to improve the understandability of the output of the tool.

%\begin{figure*}[!htb]
%\centering{\includegraphics[width=17cm]{FigureEfficiency.png}}
%\caption{Efficiency results.}
%\label{fig:efficiency}
%\end{figure*h}

In general, the evaluators agree that the tool achieves its objectives and that it allows the definition of all the security threats and the security controls of a multi-cloud application. Table \ref{tab:1} presents some figures regarding efficiency collected during the evaluation. The timing results indicated by the evaluation teams are aligned with the values obtained in internal tests performed in continuous evaluation by the tool developers. Scenario 2 was more complex, as we can conclude seeing that times are larger and that the number of controls supported was lower. As a general rule, the estimated time saved by using the tool is a very complex question. The evaluators could not give any estimation on that, because many of them did not perform any risk assessment before, although they do agree and recognize the benefit of using the tool. More experiments should be performed in the future to actually compare time using a traditional risk assessment tool, to be able to provide a more accurate answer.

\begin{table*}[t]
  \centering
    \caption{Efficiency questions related to risk assessment of a multi-cloud application.}
    \begin{tabular}{ccccc} 
    \toprule
       \multirow{2}{*}{Efficiency questions}& \multicolumn{2}{c}{Scenario1} & \multicolumn{2}{c}{Scenario2}\\        
        \cmidrule{2-5}
         & Avg. & Median & Avg. & Median\\
        \midrule
        Time spent defining the risks of one component (minutes) & 23  & 30 & 24.8 & 26\\ 
        \midrule
        Time spent defining the risks of the whole application (minutes)  & 82 & 90 & 143 & 90\\ 
        \midrule
        Estimated time saved (by using the tool) defining the risks of the application  (minutes) & 185 & 60 & 247,5 & 195 \\
        \midrule
        Number of required security controls that could be specified in the risk analysis & 67 & 86 & 36.5 & 11.5 \\
        \bottomrule
    \end{tabular}

  \label{tab:1}
\end{table*}

%\begin{figure*}[!htb]
%\centering{\includegraphics[width=17cm]{FigureUsability.png}}
%\caption{Usability results.}
%\label{fig:usability}
%\end{figure*}

%\subsection{Flexibility}

%Regarding flexibility, Figure \ref{fig:flexibility} shows how the evaluators were not sure about how well the tool scales when the number of components of the application increases or when the number of security requirements increases. This has a direct impact in our challenge around continuous risk assessment (C2). This challenge entails, among other things, being able to support an efficient risk assessment when the multi-cloud application starts to scale.

%\begin{figure*}[!htb]
%\centering{\includegraphics[width=17cm]{FigureFlexibility.png}}
%\caption{Flexibility results.}
%\label{fig:flexibility}
%\end{figure*}

%\subsection{Other evaluation results}

Finally, it is worth mentioning that some evaluators commented that perhaps the risk analysis should be done before modelling the application. Our methodology requires an initial model to offer a first analysis of risks but, obviously, it supports iterating as many times as possible between model definition and risk analysis as needed.

\subsection{Risk analysis relevance}

Finally, in order to understand the relevance of managing risks when creating a multi-cloud application, we asked evaluators to rank our risk management tool with respect to other tools to build and secure multi-cloud applications, both in terms of importance and innovation. This list of tools is based on the tools proposed in \cite{Rios2015TowardsSM}, where the authors propose a framework to support the security-intelligent lifecycle management of distributed applications over heterogeneous cloud resources. After gathering the results, the ordered list of tools was:
\begin{enumerate}
  \item a risk analysis tool
    \item a security assurance platform for monitoring
    \item a decision support system for Cloud Service Selection
    \item an Service Level Agreement (SLA) Generator
    \item a deployer to support distributed deployment
\end{enumerate}

As we can see, the risk analysis tool is the one valued highest, where 4 out of the 9 evaluators ranked it as their first choice.

\section{Conclusions}\label{sec5}
\label{Sec_Conclusions}

%There are challenges in taking agile methodologies to enterprise scale and the management of risk is one of the neglected challenges%, frequently assuming significance when an application is deployed.

Risk assessment is often an afterthought, as it happens with security as a whole. Risk assessment is usually performed in a quick and unstructured manner or even completely skipped. As a consequence, risk assessment often leads to ineffective and inaccurate analysis.

Considering risk management properly in the agile development process generates a number of challenges to be solved. % from the introduction of a top-down non functional requirement into a bottom-up functionally based methodology. 
 This paper combines the information collected from previous works together with years of internal experience to describe several pending challenges for risk management. %, that, if tackled correctly, can increase the quality of the developed application.
%All of the research challenges will move the state of the art forward.
The proposed challenges also cover one area that is often neglected: team cultural change. By creating tools that support the aforementioned challenges we will enable faster and more comprehensive adoption of agile risk management tools and techniques.

In order to cover the challenges identified in this paper, we have proposed a framework that is based on an online Kanban-like tool that is agile (challenge C1) and fosters collaboration (challenge C4) by offering a visual representation of the proposed risks/threats and their related mitigation actions. This framework proposes using recommendations and rules to offer automation and guidance to the team. This makes risk assessment attainable and usable even by software designers that have good technical skills but may not be security and risk analysis experts (challenge C3). Moreover, since the knowledge database can be tailored to the specific domain of the application, a finer granularity in the risks and mitigation actions suggested can also be achieved. Our proposal allows the natural introduction of new risks and threats and re-evaluation of the level of security related to each system component following continuous software delivery methods (challenge C2).

In this paper, we have also described an implementation of the framework in order to perform a user evaluation. %The selected methodology simplifies the security-by-design approach to risk assessment to 4 main steps: risk selection, risk evaluation (defining their likelihood and impact values), selection of the mitigation actions that the tool suggests, and evaluating the status of the risks.  
From the results of this evaluation we can conclude that an agile risk analysis tool is one of the main tools needed to develop a secure multi-cloud application. Moreover, our tool and selected methodology received a very positive feedback from the evaluations and were able to satisfy the needs to tackle all challenges. 

For future work, apart from some GUI glitches, for future versions we would like to improve the scalability of the tool by offering operations that can affect multiple components at the same time or that can allow the user to add or remove groups of security controls in one operation. We would then like to perform a new evaluation in order to assess the impact of our changes and to better measure how this methodology improves past techniques and how cultural change can be pushed via an agile tool. Another important line of future work is adding automation, so that the tool can learn from the users' past behaviour and proactively suggest actions or add risks for components similar to those that the user has analysed in the past. In this sense, developing a tool that could automatically suggest risks from the definition of the application components would boost the impact on teams with little risks-related experience. From a different perspective, we also plan to apply this methodology and develop a similar tool to handle risks associated to software development planning.

\section*{Acknowledgements}
  %The authors would also like to thank the anonymous referees for their valuable comments and helpful suggestions.
  This work is supported by the European Commission through the \emph{ENACT} project under Project ID:~780351 and the \emph{PDP4E} project under Project ID:~787034.

\bibliographystyle{plain}
\bibliography{agileRisk} 

\end{document}